# Numerical solution of the Schrödinger equation for types of Woods-Saxon potential

**Mirzaei Mahmoud Abadi. Vahid[a*]**

**Hosseini Ranjbar. Abbas[a]**

**Javad. Mohammadi[a]**

**Rahim Khabaz Kharame[b]**

[a] Faculty of Physics, Shahid Bahonar University, P.O. Box 76175, Kerman, Iran

## Abstract

In this study, the Schrödinger equation for the Woods-Saxon potential, the general Woods-Saxon potential, and D-dimensional Woods-Saxon potential is numerically investigated. The finite difference method is used to transform the Schrödinger equation to an eigensystem and the Jacobi method is applied to calculate eigenpairs. It has been noted that Schrödinger equation with Woods-Saxon potential expect for $l$=0 does not have the exact solution. Some approximate analytical solutions for this type of potential is pointed out. Finally, the obtained energy levels are compared to the similar studies in special cases for $^{56}Fe$. Normalized wave functions are calculated and for some cases are graphically shown.

## Introduction

The exact solution of the Schrödinger equation in physical problems is crucial as it gives a deep understanding of the physics problem. There are several methods for solving this equation, which provide valuable solutions to the problem of physical models [1-4].

Solving this equation for radius-centered potentials in recent years has attracted many researchers, such as the Rosen-Morse potential, the Morse potential, the Yukawa potential, and etc.

For some potentials, there is no analytical solution to the Schrödinger equation. In this case, methods based on the theory of perturbation such as the WKB method, approximate and numerical methods such as factorization [4], SASYQM APPROACH [5-7], semi-linearization [8,9], Nikiforov-Uvarov [10] is used.

In the perturbation methods, the perturbed potential should be small in comparison to the Hamiltonian of the system. One of the problems with perturbation methods is the need to solve a series of complex integrals. The use of numerical methods is appropriate because of the fact that there is no specific limit to the type of potential. For this reason, the numerical solution of the Schrödinger equation is of interest to researchers.

As was mentioned the numerical solution of the Schrödinger equation leads to an eigensystem. Necessary boundary conditions for transforming the Schrödinger equation to an eigensystem must be imposed at initial and endpoint. The wave function must become zero at these two points. There are some methods for solving eigensystems. Each method has some strengths and weaknesses. Power method, the reduction method, the Givens method, and the Jacobi method are some examples of numerical methods for solving eigensystems. In Jacobi transformation method a similarity transform is used to diagonal coefficients matrix. The matrix is reduced down to a form with little two-by-two blocks step by step by several iterations. Suppose $A$ is the coefficients matrix then

$$A \to Z^{-1} \cdot A \cdot Z \qquad (1)$$

is called similarity transform. The procedure in Jacobi transformation method is

$$A \to P_1^{-1} \cdot A \cdot P_1 \to P_2^{-1} \cdot P_1^{-1} \cdot A \cdot P_1 \cdot P_2$$

$$\to P_3^{-1} \cdot P_2^{-1} \cdot P_1^{-1} \cdot A \cdot P_1 \cdot P_2 \cdot P_3 \to \quad etc. \qquad (2)$$

Suitable selecting of $P_i$s after several iterations $A$ become diagonal. By introducing $X_R$ in the form

$$X_R = P_1 \cdot P_2 \cdot P_3 \cdot \cdots \qquad (3)$$

it is proved that $X_R$ columns are eigenvectors of $A$ and transformed $A$ will be diagonal in which its diagonal elements are the eigenvalues of the eigensystem. The number of mesh which is the order of eigensystem depends on the accuracy of the solution. Jacobi method gives the same number of eigenvalues and eigenvectors as the number of mesh. It is seen that the eigenvalues accuracy decreases with increasing the absolute of eigenvalues.

A potential that has been taken into consideration over the years is Woods-Saxon potential. This potential was proposed in 1954 by the scientists (R. D. Woods) and (D. S. Saxon) to explain the elastic scattering dispersion of protons of 20 MeV through medium and heavy nuclei [11]. By subsequent studies, empirical observations, and solving Schrödinger equation with this potential a useful model for determining the particle energy level of nuclei, named the shell model was proposed to provide energy levels accurately [12]. This potential also examines the interactions within the nucleus, such as nucleon-nucleon interactions. In microscopic physics, it described a complete description of the interaction of a nucleus with a heavy nucleus. It also acts as a central part in the interaction of neutrons with a heavy nucleus [13-15]. The deformed Woods-Saxon potential is also a short-range potential which is widely used in several fields of sciences such as nuclear, particles, atomic, condense matter and physical chemistry [16-18]. The Woods-Saxon potential has likewise been used for the description of heavy ions elastic scattering at low energies [19]. In this study, the Schrödinger equation with Woods-Saxon potential, general Woods-Saxon potential, and D-dimensional Woods-Saxon potential has been solved numerically by Jacobi method.

The standard form of this potential is:

$$V(r) = \frac{v_0}{1+e^{\frac{r-R_0}{a_0}}} \qquad (4)$$

where $v_0$, $R_0$ and $a_0$ are the potential depth, the potential width, and the thickness of the surface respectively.

The general version of this potential is:

$$V(r) = -\frac{v_0}{1+e^{\frac{r-R_0}{a_0}}} - \frac{w_0 e^{\frac{r-R_0}{a_0}}}{(1+e^{\frac{r-R_0}{a_0}})^2} \qquad (5)$$

It consists of an additional part, which is the derivative of the standard form of this potential. This type of potential is well-known to generalized Wood-Saxon potential. The shell model in nuclear physics is made by adding the interaction of *l.s* to this potential [20-22].

One particular point regarding the general Woods-Saxon is that for $w_0 < 0$, there is a potential barrier that is used to describe the quasi-bound states in the nucleus. Moreover, if $R_0 = 0$ the potential is changed to the Rosen-Morse potential [2, 4].

The effective form of potential will be

$$V_{eff}(\mathrm{r}) = -\frac{v_0}{1+e^{\frac{r-R_0}{a_0}}} - \frac{w_0 e^{\frac{r-R_0}{a_0}}}{\left(1+e^{\frac{r-R_0}{a_0}}\right)^2} + \frac{l(l+1)\hbar^2}{2\mu r^2} \qquad (6)$$

where $\mu$ is the reduced mass. The NU method is one of the methods which is able to use for several potentials.

Radial part of Schrödinger equation for D-dimensional case has been presented in [23, 24]

$$\left[\frac{d^2}{dr^2} + \frac{D-1}{r}\frac{d}{dr} + \frac{2\mu}{\hbar^2}\left(E_{n,l}^D + V(r)\right) - \frac{l(l+D-2)}{r^2}\right] R_{n,l}(r) = 0, \quad 0 \le r \le \infty; \qquad (7)$$

In the D-dimensional case for removing first order derivative, the well-known transformation is used.

In the next sections Woods-Saxon, general Woods-Saxon, and D-dimensional Woods-Saxon potentials are numerically investigated. In the present study using finite difference method, Schrödinger equation transforms into an eigensystem matrix. Then the eigensystem is solved by the Jacobi method. Finally, the eigenvalues which are energies of the eigensystem and the eigenvectors which are related to the wave functions are obtained. Our results are given in figures and tables. The results are compared with other researches.

## The Woods-Saxon Potential

Using the well-known transforming statement

$$U_{n,l}(r) = rR_{n,l}(r) \qquad (8)$$

the differential equation of the radial part of Schrödinger equation with Woods-Saxon potential is given by

$$\frac{d^2U_{n,l}(r)}{dr^2} + \frac{2\mu}{\hbar^2}\left[E_{n,l} + \frac{V_0}{1+e^{\frac{r-R_0}{a}}} - \frac{\hbar^2 l(l+1)}{2\mu r^2}\right] U_{n,l}(r) = 0 \qquad (9)$$

where $\mu = m_A m_n/(m_A + m_n)$. For $l$=0, equation (9) has the analytical solution [25]. But, the Woods-Saxon potential does not have the exact solution for $l \neq 0$. In this case in which there is not an exact analytical solution for Woods-Saxon potential an approximation is applied as follow [26]

$$\frac{1}{r^2} \cong \frac{\delta^2 e^{\delta r}}{(1-e^{\delta r})^2} \qquad (10)$$

where

$$\delta = \frac{l(l+1)\hbar^2}{2\mu r^2} \qquad (11)$$

Employing this approximation by the NU method, the Schrödinger equation transforms into a hypergeometric equation with a series of transformations, that its solution is known [10]. Authors have driven the energy eigenvalues and wave functions of Woods-Saxon, (32), (39) of reference [27].

The energy eigenvalues and some normalized wave functions of the bound state of Woods-Saxon potential for a neutron orbiting around $^{56}Fe$ are given in figure 1. Potential parameters are the same as [22], $r_0 = 1.285$ fm, $a = 0.65$ fm, $V_0 = (40.5 + 0.13A)$ MeV $= 47.78$ MeV, $R_0 = r_0 A^{1/3} = 4.9162$ fm, $m_{\text{core}} = 56u$, $m_n = 1.00866u$. All normalized wave functions up to $l$=3 are shown.

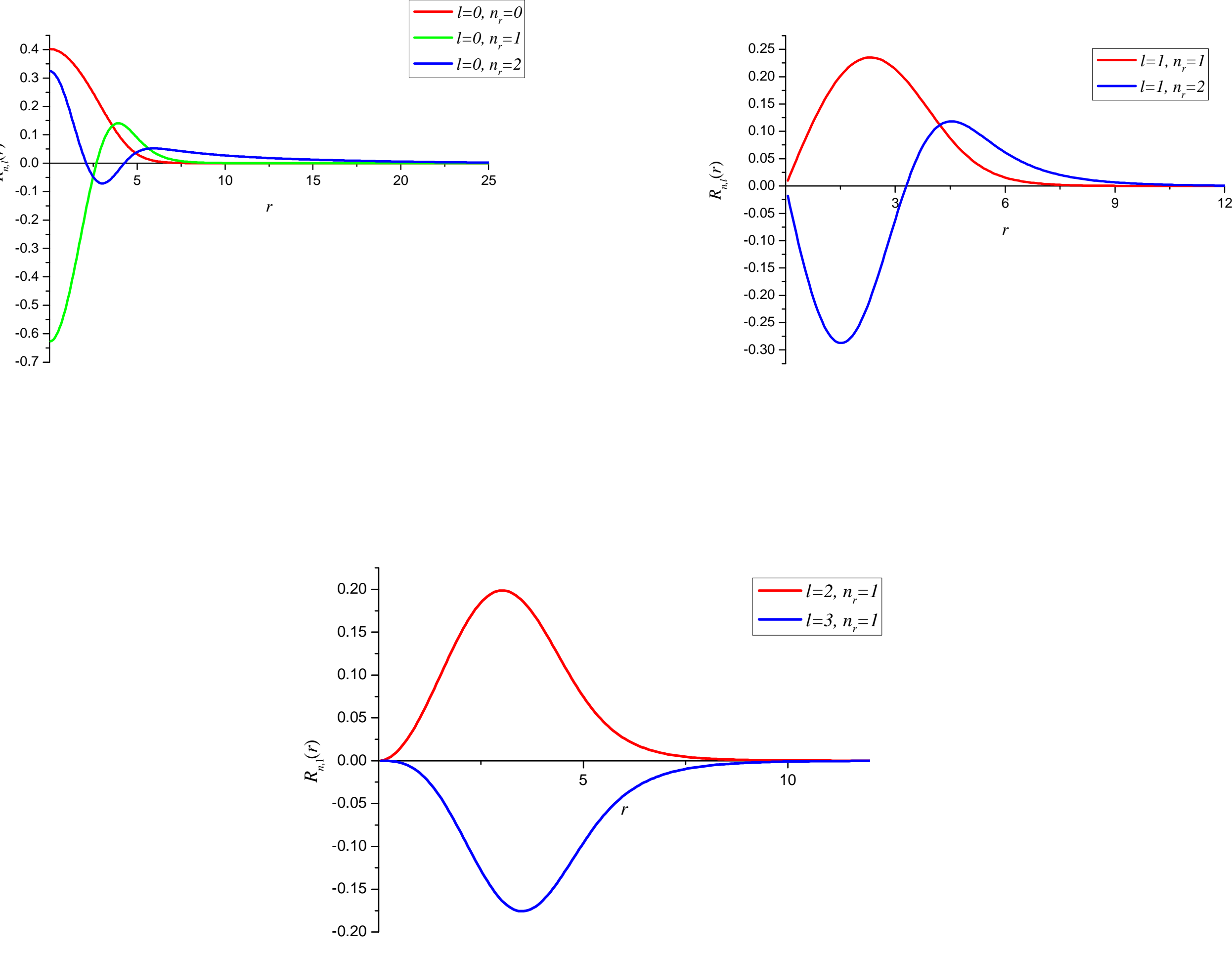

**Figure 1.** Normalized wave functions of Woods-Saxon potential for a neutron interaction with $^{56}Fe$.

It must be noted that obtained eigenvectors by Jacobi method are normalized, but, corresponding eigenfunctions are not normalized. Therefore, a simple relation that should exist between eigenvectors and corresponding wave functions, the normalized eigenfunctions are obtained.

$$\int_{x_{min}}^{x_{max}} |u(x)|^2 dx \rightarrow h \sum_{i=0}^{N} u_i^2 = 1, \;\; h = \frac{x_{max} - x_{min}}{N}$$

The Jacobi method is a suitable method for eigenvalue problems in physics. Thus, our FORTRAN program using Jacobi method should numerically solves the eigensystems. We should, certainly, evaluate our program for several potentials that have analytical solution. This program is checked by several potentials which have the analytical solution. As is expected, increasing $n_r$ the wave function broadening occurs at larger $r$.

## The General Woods-Saxon Potential

The general form of Woods-Saxon potential is presented in equation (5). Radial part of the Schrödinger equation for general Woods-Saxon potential is

$$\frac{d^2U_{n,l}(r)}{dr^2}+\frac{2\mu}{\hbar^2}\left[E_{n,l}+\frac{V_0}{1+e^{\frac{r-R_0}{a}}}+\frac{w_0 e^{\frac{r-R_0}{a_0}}}{(1+e^{\frac{r-R_0}{a_0}})^2}-\frac{\hbar^2 l(l+1)}{2\mu r^2}\right]U_{n,l}(r)=0 \qquad (12)$$

The approximate analytical solution for general Woods-Saxon potential is presented by the authors. The energy eigenvalues are obtained from equation (13) of reference [26], where the employed approximate to solve the Schrödinger equation is expressed in equation (10). The UN method have been applied to solve the corresponding radial part of the Schrödinger equation (12). The unnormalized wave functions are given by equation (14) in reference [26]. The results of our calculations for some $l$ and $W_0$ are given in table 1. Some normalized wave functions are also shown in figure 2. To compare the results, calculations were performed for a neutron interaction with $^{56}Fe$. The potential parameters are the same as before.

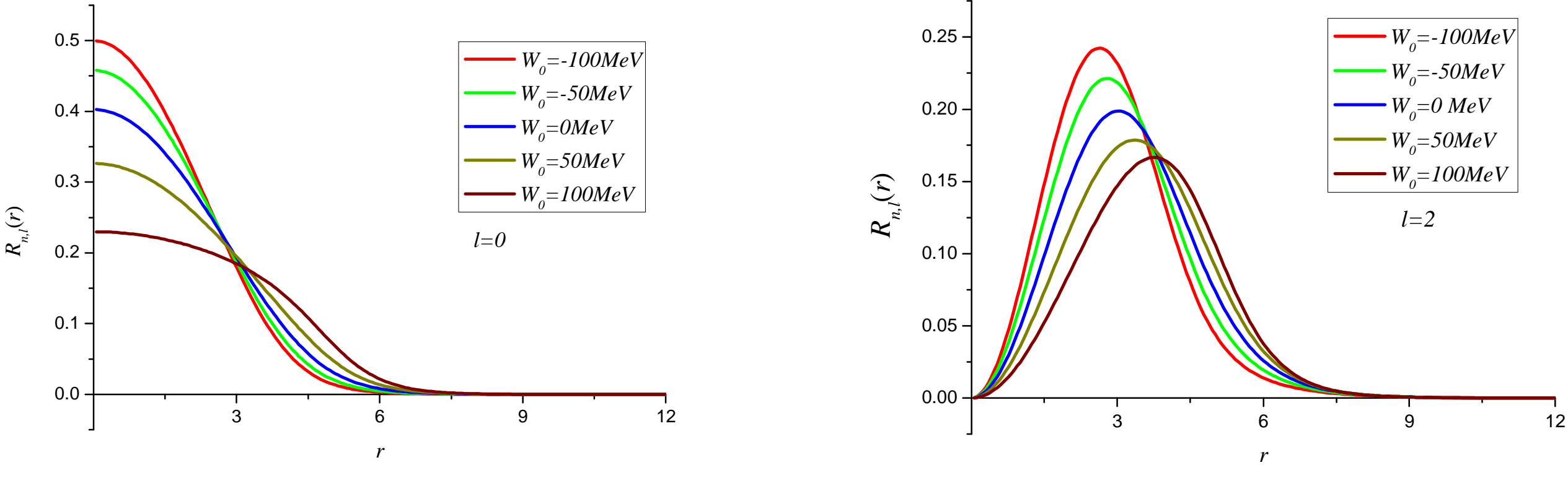

**Figure 2.** Normalized wave functions of general Woods-Saxon potential for various values of $W_0$ from -100MeV to 100MeV with 50MeV step length for a neutron interacting with $^{56}Fe$.

The results show that for the small values at the $W_0$ the sharpness of the wave function increases and vice versa.

## The D-dimensional Woods-Saxon and D-dimensional general Woods-Saxon Potential

D-dimensional Woods-Saxon potential has been studied by several authors [28, 29]. The radial part of the Schrödinger equation for the D-dimensional Woods-Saxon potential is in the form (7). Removing the first order derivative leads to the following well-known transforming equation

$$U_{n,l}(r) = r^{\frac{D-1}{2}} R_{n,l}(r) \qquad (13)$$

**Table 1.** Comparison of the single particle energy levels of neutron interaction with $^{56}Fe$ nucleus employing several potential depth of $W_0$ and quantum numbers of $n_r$ for $l$= 0. The potential parameters are given in the text.

| $W_0(MeV)$ | $n_r$ | $E_{nr}$(Ref[26]) | $E_{nr}$(Ref[31]) | $E_{nr}$(present study) |
|---|---|---|---|---|
| 0 | 0 | -38.3004 | -38.3002 | -38.2930 |
| 0 | 1 | -18.2254 | -18.2227 | -18.2018 |
| 0 | 2 | -0.2678 | -0.2663 | -0.2556 |
| 0 | 3 | 62.9775 | unbound | unbound |
| 50 | 0 | -41.1965 | -41.1964 | -41.1893 |
| 50 | 1 | -23.8789 | -23.8788 | -23.8550 |
| 50 | 2 | -3.6472 | -3.6471 | -3.6181 |
| 50 | 3 | 52.0232 | unbound | unbound |
| 100 | 0 | -45.4453 | -45.4446 | -45.4384 |
| 100 | 1 | -29.1659 | -29.1642 | -29.1410 |
| 100 | 2 | undetermined | -7.8143 | -7.7787 |
| 100 | 3 | undetermined | unbound | unbound |
| -50 | 0 | -36.2136 | -36.2168 | -36.2065 |
| -50 | 1 | -12.8469 | -12.8504 | -12.8223 |

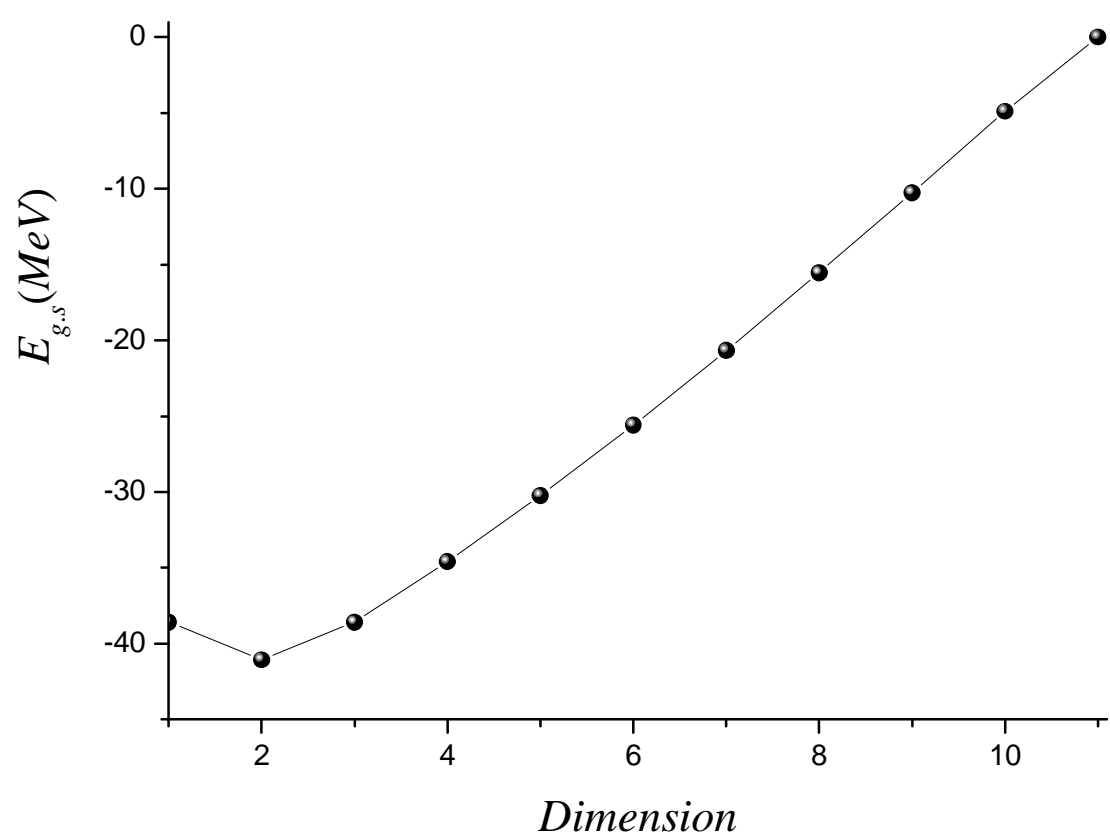

**Figure 3.** Ground state energy of Woods-Saxon potential versus dimension the Schrödinger equation for neutron interaction with $^{56}Fe$. The potential parameters are given in the text.

Applying this transformation gives:

$$\frac{d^2U_{n,l}(r)}{dr^2}+\frac{2\mu}{\hbar^2}\left[E_{n,l}^{D}+\frac{V_0}{1+e^{\frac{r-R_0}{a}}}-\frac{\hbar^2(D+2l-1)(D+2l-3)}{8\mu r^2}\right]U_{n,l}(r)=0 \qquad (14)$$

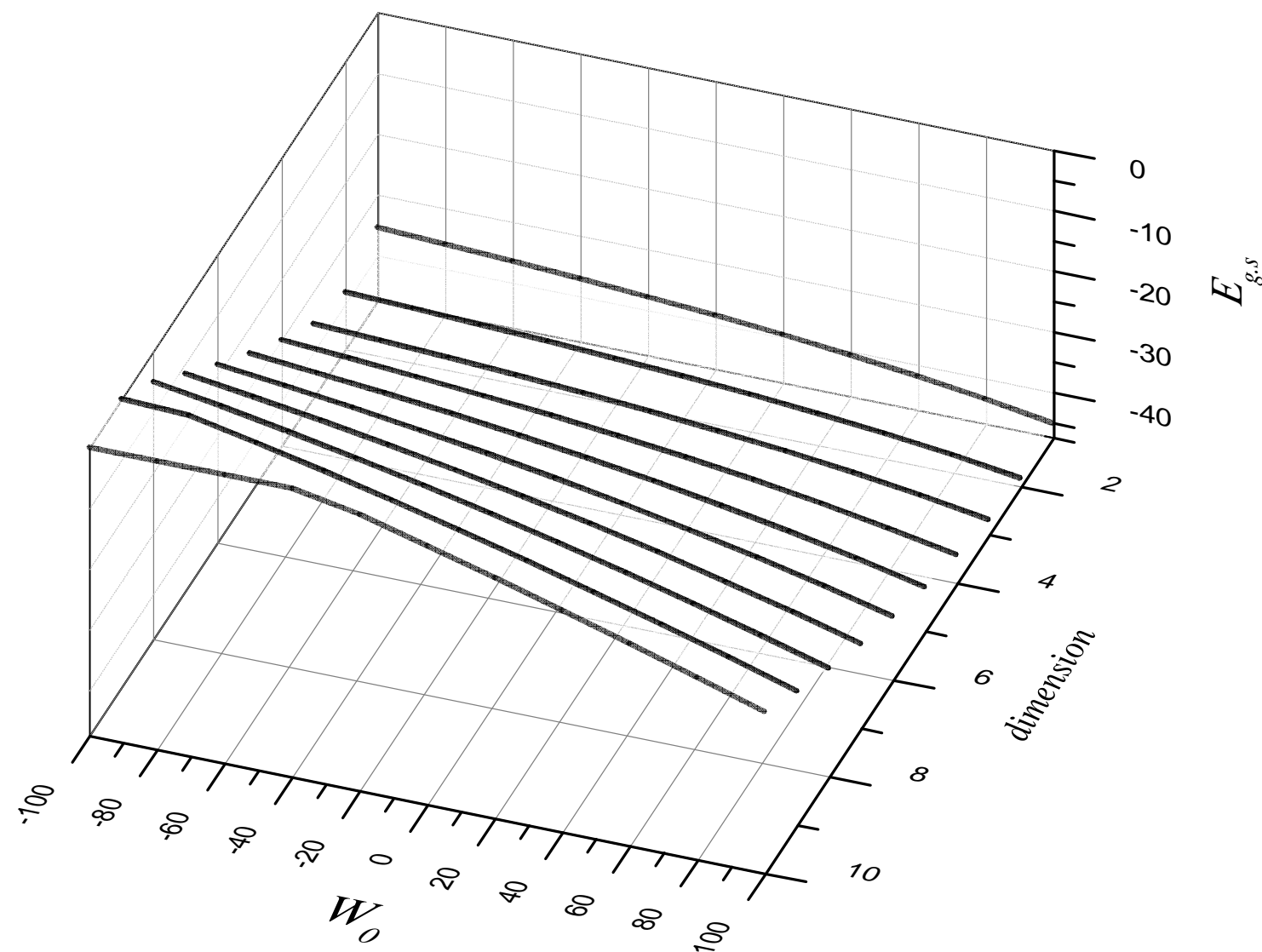

**Figure 4.** A 3D-curve of ground state energy versus dimension and $W_0$ for general Woods-Saxon potential. The curve is plotted for $^{56}Fe$ and the potential parameters are given in the text.

For general Woods-Saxon potential we will have

$$\frac{d^2U_{n,l}(r)}{dr^2}+\frac{2\mu}{\hbar^2}\left[E_{n,l}^{D}+\frac{V_0}{1+e^{\frac{r-R_0}{a}}}+\frac{w_0 e^{\frac{r-R_0}{a_0}}}{(1+e^{\frac{r-R_0}{a_0}})^2}-\frac{\hbar^2(D+2l-1)(D+2l-3)}{8\mu r^2}\right]U_{n,l}(r)=0 \quad (15)$$

Schrödinger equation with the D-dimensional general Woods-Saxon potential has been solved by several authors. The energy eigenvalues are presented by V. H. Badalov et al in equation (3.17) of reference [30].

## Conclusion

According to our calculation the Jacobi method which we employed for several potentials is a suitable method for physical problems. Numerical methods have an important role in physics and quantum mechanics, since, often there is no analytical solution for the Schrödinger equation. Hence, numerical solutions could help us to find the deep understanding of a problem. Jacobi method gives more accurate eigenvalues for smaller ones in the spectrum of the eigenvalues. This is a good approach in physics problems as in most cases the lowest energy levels are of great significance. There are several positive energy eigenvalues for hydrogen atom and the Woods-Saxon potential. These energy eigenvalues correspond to quasi-bound states and states which are related to the scattering. Therefore, the Jacobi method could, also be used for scattering problems.

## References

[1] Ikot, A. N., L. E. Akpabio, and E. B. Umoren. "Exact solution of Schrödinger equation with inverted woods-saxon and Manning-Rosen potentials." *Journal of Scientific Research* 3.1 (2010): 25.

[2] Cooper, Fred, Avinash Khare, and Uday Sukhatme. "Supersymmetry and quantum mechanics." *Physics Reports* 251.5-6 (1995): 267-385

[3] Morales, Daniel A. "Supersymmetric improvement of the Pekeris approximation for the rotating Morse potential." *Chemical physics letters* 394.1-3 (2004): 68-75.

[4] Flügge, Siegfried. *Practical quantum mechanics*. Springer Science & Business Media, 2012.

[5] Gönül, B. "Exact treatment of $\ell\neq 0$ states." *Chin. Phys. Lett.* 21.quant-ph/0407004 (2004): 1685.

[6] Dutt, R., K. Chowdhury, and Y. P. Varshni. "An improved calculation for screened Coulomb potentials in Rayleigh-Schrodinger perturbation theory." *Journal of Physics A: Mathematical and General* 18.9 (1985): 1379.

[7] Filho, Elso Drigo, and Regina Maria Ricotta. "Supersymmetry, variational method and Hulthen potential." *Modern Physics Letters A* 10.22 (1995): 1613-1618.

[8] Mandelzweig, V. B. "Comparison of quasilinear and WKB approximations." *Annals of Physics* 321.12 (2006): 2810-2829.

[9] Ikhdair, Sameer M., and Ramazan Sever. "Approximate eigenvalue and eigenfunction solutions for the generalized Hulthén potential with any angular momentum." *Journal of Mathematical Chemistry* 42.3 (2007): 461-471.

[10] Nikiforov, Arnold F., and Vasilii B. Uvarov. *Special functions of mathematical physics*. Vol. 205. Basel: Birkhäuser, 1988.

[11] Woods, Roger D., and David S. Saxon. "Diffuse surface optical model for nucleon-nuclei scattering." *Physical Review* 95.2 (1954): 577.

[12] Nicolai, Hermann. "Supersymmetry and spin systems." *Journal of Physics A: Mathematical and General* 9.9 (1976): 1497.

[13] Bohr, Aage, and Ben R. Mottelson. *NUCLEAR STRUCTURE (IN 2 VOLUMES):(In 2 Volumes) Volume I: Single-Particle MotionVolume II: Nuclear Deformations*. World Scientific Publishing Company, 1998.

[14] Brandan, María-Ester, and George R. Satchler. "The interaction between light heavy-ions and what it tells us." *Physics reports* 285.4-5 (1997): 143-243.

[15] Satchler, George Raymond. "Heavy-ion scattering and reactions near the Coulomb barrier and "threshold anomalies"." *Physics Reports* 199.3 (1991): 147-190.

[16] Garcia, F., et al. "Woods-Saxon potential parametrization at large deformations for plutonium odd isotopes." *The European Physical Journal A-Hadrons and Nuclei* 6.1 (1999): 49-58.

[17] Diaz-Torres, Alexis, and Werner Scheid. "Two center shell model with Woods–Saxon potentials: adiabatic and diabatic states in fusion." *Nuclear Physics A* 757.3-4 (2005): 373-389.

[18] Guo, Jian-You, and Zong-Qiang Sheng. "Solution of the Dirac equation for the Woods–Saxon potential with spin and pseudospin symmetry." *Physics Letters A* 338.2 (2005): 90-96.

[19] Bespalova, O. V., E. A. Romanovsky, and T. I. Spasskaya. "Nucleon–nucleus real potential of Woods–Saxon shape between– 60 and+ 60 MeV for the $40 \leqslant A \leqslant 208$ nuclei." *Journal of Physics G: Nuclear and Particle Physics* 29.6 (2003): 1193.

[20] Zaichenko, A. K., and V. S. Ol'Khovskii. "Analytic solutions of the problem of scattering by potentials of the Eckart class." *Theoretical and Mathematical Physics* 27.2 (1976): 475-477.

[21] Fakhri, H., and J. Sadeghi. "Supersymmetry approaches to the bound states of the generalized Woods–Saxon potential." *Modern Physics Letters A* 19.08 (2004): 615-625.

[22] Berkdemir, Cüneyt, Ayşe Berkdemir, and Ramazan Sever. "Polynomial solutions of the Schrödinger equation for the generalized Woods-Saxon potential." *Physical Review C* 72.2 (2005): 027001.

[25] Aktas, Metin, and Ramazan Sever. "Exact supersymmetric solution of Schrodinger equation for central confining potentials by using the Nikiforov-Uvarov method." *arXiv preprint hep-th/0409139* (2004).

[26] Bayrak, O., and E. Aciksoz. "Corrected analytical solution of the generalized Woods–Saxon potential for arbitrary states." *Physica Scripta* 90.1 (2014): 015302.

[27] Ikot, Akpan N., and Ita O. Akpan. "Bound State Solutions of the Schrödinger Equation for a More General Woods—Saxon Potential with Arbitrary l-State." *Chinese Physics Letters* 29.9 (2012): 090302.

[28] Niknam, A., A. A. Rajabi, and M. Solaimani. "Solutions of D-dimensional Schrodinger equation for Woods–Saxon potential with spin–orbit, coulomb and centrifugal terms through a new hybrid numerical fitting Nikiforov–Uvarov method." *Journal of Theoretical and Applied Physics* 10.1 (2016): 53-59.

[29] Badalov, V. H., and H. I. Ahmadov. "Analytical solutions of the $ D $-dimensional Schr\"{o} dinger equation with the Woods-Saxon potential for arbitrary $ l $ state." *arXiv preprint arXiv:1111.4734* (2011).

[30] arXiv:1711.10322[nucl-th]

[31] Vertse, T., K. F. Pál, and Z. Balogh. "GAMOW, a program for calculating the resonant state solution of the Radial Schrödinger Equation in an arbitrary optical potential." *Computer Physics Communications* 27 (1982): 309-322.